# Electronic structure and glass forming ability in early and late transition metal alloys

E. Babić[a], R. Ristić[b]*, I. A. Figueroa[c], D. Pajić[a], Ž. Skoko[a] and K. Zadro[a]

[a]Department of Physics, Faculty of Science, University of Zagreb, Zagreb, Croatia; [b]Department of Physics, University of Osijek, Osijek, Croatia; [c]Institute for Materials Research-UNAM, Ciudad Universitaria Coyoacan, Mexico, Mexico

*corresponding author: Ramir Ristić, email:ramir.ristic@fizika.unios.hr

A correlation between the change in magnetic susceptibility ($\Delta\chi_{exp}$) upon crystallization of Cu-Zr and Hf metallic glasses (MG) with their glass forming ability (GFA) observed recently, is found to apply to Cu-Ti and Zr-Ni alloys, too. In particular, small $\Delta\chi_{exp}$, which reflects similar electronic structures, ES, of glassy and corresponding crystalline alloys, corresponds to high GFA. Here, we studied $\Delta\chi_{exp}$ for five Cu-Ti and four Cu-Zr and Ni-Zr MGs. The fully crystalline final state of all alloys was verified from X-ray diffraction patterns. The variation of GFA with composition in Cu-Ti, Cu-Zr and Cu-Hf MGs was established from the variation of the corresponding critical casting thickness, $d_c$. Due to the absence of data for $d_c$ in Ni-Zr MGs their GFA was described using empirical criteria, such as the reduced glass transition temperature. A very good correlation between $\Delta\chi_{exp}$ and $d_c$ (and/or other criteria for GFA) was observed for all alloys studied. The correlation between the ES and GFA showed up best for Cu-Zr and $NiZr_2$ alloys where direct data for the change in ES ($\Delta ES$) upon crystallization are available. The applicability of the $\Delta\chi_{exp}$ ($\Delta ES$) criterion for high GFA (which provides a simple way to select the compositions with high GFA) to other metal-metal MGs (including ternary and multicomponent bulk MGs) is briefly discussed.

1. **Introduction**

It is well known , for example ref. [1], that detailed insight into the formation of the amorphous state is necessary for future understanding and applications of both insulating [2] and metallic glasses (MG) [3, 4]. While first-principles understanding of the amorphous state is still an open problem [5, 6] a key issue for the application of MGs is understanding the glass forming ability (GFA), i.e. how the critical cooling rate necessary for vitrification $R_c$, or equivalently the maximum casting thickness $d_c$, depends on the components and composition of the alloy [3, 4, 7, 8]. As result, a search for alloy parameters correlating with GFA [9] started simultaneously with the discovery of MGs [10] and has accelerated upon the discoveries of bulk metallic glasses (BMG) with $d_c \geq 10$ mm [11, 12] which are promising as structural and functional materials [4, 8] with technological applications.

These empirical or semi-empirical criteria for high GFA are mostly based on thermodynamic parameters [3, 4, 9], characteristic temperatures (such as the reduced glass transition temperature $T_{rg}$ [13] and other similarly constructed parameters [14, 15]), as well as enthalpies [16] and free energies or entropies [13, 17]. The atomic size mismatch, which destabilizes the crystalline lattice [15, 18] and the effective valence $Z_{eff}$, which is expected to stabilize the amorphous phase [19], were also proposed to correlate with GFA. We note however that there is no evidence for a correlation between $Z_{eff}$ and GFA [20, 21]. Since the most of these criteria were designed for binary systems, they work quite well for some binary and ternary alloy systems (e.g., [16, 18, 20, 21]), but perform less well for the more important multicomponent high-$d_c$ BMGs [14, 15, 22]. Among the novel criteria intended to explain the high GFA in BMGs, Inoue´s rules [18] combining the „confusion" principle, strong chemical interaction, atomic size mismatch and high packing density seem most complete. Some other criteria evoke modest chemical interactions, indicated by volume conservation, and frustration due to competing crystalline phases (CPs) [20, 23], the fragility of the undercooled melt [4, 22, 24], etc. Further, numerical simulations are used in order to associate GFA with efficient packing of atomic clusters in MGs [25], or to test the relative importance of the various factors entering into Inoue´s rules [26]. Recent experiments with amorphous high entropy alloys a-HEA , for example [27], seemed to further complicate the problem of the origin of GFA, but more careful analysis [15, 28] indicates that the stability and GFA of a-HEA alloys containing the early transition metals (TE) behave similarly to those in conventional MGs containing TEs [20, 21].

Recently we noted [20, 29] that to our knowledge no criterion for GFA in metallic alloys takes their electronic band structure (ES) explicitly into account, in spite of the fact that in metallic systems, a large contribution to the cohesive energy comes from the conduction electrons, which makes their properties very sensitive to their ES. Accordingly, we proposed that similar ES of MG and the corresponding crystalline alloy enhances GFA [20, 29]. This proposal follows from a well known fact that high GFA results from similar free energies of MG and competing/primary CP(s) [3, 30]. At low temperatures , T < $T_g$, the glass transition temperature, the free energy is dominated by the internal energy U, and U in turn reflects the ES of metallic systems.Therefore similar ES in MG and the corresponding CP(s) is clearly important for good GFA [3, 20, 29]. This probably shows up the best for the $Zr_2Ni$ composition [31] in the Zr-Ni alloy system. This composition corresponds to a local maximum of GFA in this alloy system as shown in Ref. [32] and also in Fig. 6, in spite of the fact that $Zr_2Ni$ is a stable intermetallic compound. At first, we tested our proposal for Cu-Hf, Zr alloys [1] by comparing the changes in the properties directly related to ES, such as the magnetic susceptibility $\chi_{exp}$, and the coefficient of a linear term in the low temperature specific heat (LTSH) γ, upon crystallization of MGs with the corresponding reduced glass transition temperatures $T_{rg}$, where $T_{rg} = T_g/T_l$ with $T_l$ and $T_g$ the liquidus and glass transition temperatures, respectively [13]. We selected the alloys of TE with the late transition metals (TL) because some data for Zr-Cu alloys did already exist [33, 34] and because in TE-TL MGs there is a rather simple correlation between their properties and ES [20, 21, 35]. Further, valence band spectra of crystalline Zr-Cu, Pd alloys are close to those of MGs of the same composition, for example P. Steiner et al in [31]. For all alloys studied, the changes in $\chi_{exp}$ and γ showed good correlation with $T_{rg}$: small changes in $\chi_{exp}$ and γ corresponded to large $T_{rg}$, thus to good GFA [1].

For the present paper we studied $\Delta\chi_{exp}$ in five Cu-Ti and four selected Cu-Zr and Ni-Zr alloys. The actual compositions of the Cu-Zr and Ni-Zr alloys were selected in order to complete and/or verify literature data [33, 34, 36]. An important, novel feature of the present study is that we used the experimental $d_c$ data in order to establish the variations of GFA with composition in Cu-Ti, Cu-Hf and Cu-Zr alloys. Due to the absence of $d_c$ data for Ni-Zr alloys we described GFA in this alloy system by using empirical criteria such as $T_{rg}$ and $\gamma_{GFA} = T_x/(T_l+T_g)$ [32] where $T_x$ is the crystallization temperature. For all four alloy systems studied good correlation between $\Delta\chi_{exp}$ and $d_c$ and/or $T_{rg}$ and $\gamma_{GFA}$ has been established. The correlation between the (small) changes in ES and GFA, showed up the best in Cu-Zr and $NiZr_2$ alloys for which data for the change in the electronic density of states at the Fermi level $N(E_F)$ upon crystallization are available [31, 33]. The probable applicability of this criterion for GFA to some other MGs is briefly discussed. We note however that a simplified representation of the actual ES by the electronic density of states (DOS) at the Fermi level, implied by the results presented here , may be specific to the alloys of early and late transition metals (e.g.[20,31,33,35]) and may not apply to all MGs.

2. **Experimental**

Six $Cu_xTi_{100-x}$ (x = 35, 50, 55, 60, 65, 70) glassy ribbons with similar cross-sections and therefore with the amorphous phases having broadly the same quenched-in disorder were prepared by melt-spinning fragments of arc-melted alloys in a pure helium atmosphere [20]. The amorphous state of as-cast ribbons was confirmed by differential scanning calorimetry (DSC) and x-ray diffraction (XRD) studies. The $Cu_xZr_{100-x}$ (x=33, 55, 70) and $Ni_{30}Zr_{70}$ glassy ribbons were prepared [20, 35] in practically the same way as the Cu-Ti ribbons and their amorphous state was also verified from XRD patterns. The method of preparation and DSC and XRD measurements on $Cu_xHf_{100-x}$ glassy ribbons were recently reported [1, 37]. The method used for $d_c$ measurements was also described in [37]. The same method was used to determine $d_c$ for $Zr_{60}Cu_{40}$ and $Zr_{30}Cu_{70}$ alloys. The magnetic susceptibility of glassy alloys ($\chi_a$) was measured with a Quantum Design SQUID-based magnetometer in a magnetic field B ≤ 5.5 T over the temperature range 5-300 K [1, 20]. The methods used to measure the magnetic susceptibility of Cu-Zr and Ni-Zr samples can be found in original papers [34, 36]. The same is true for LTSH measurements on Cu-Zr alloys which are described in some detail in [33]. All samples used for measurements of $\chi_a$ were later crystallized following the procedure described in [1] which is similar to that used previously for crystallization of Cu-Zr [34] and Ni-Zr [36] MGs. In particular, all alloys were heated at 10 K/min in a high-purity Ar atmosphere up to predetermined $T_a$ which corresponded to the end of the first crystallization maximum in the DSC trace of a given alloy. After a short dwell time (5-10 min) at $T_a$ the samples were furnace cooled. The annealed samples showed the same metallic shine and colour as the as-cast samples, which probably indicates a very small amount of oxidation at the surface. Such procedures were followed in order to obtain the primary crystallized samples [1], i.e. to avoid the eventual transformation of primary CP(s). All the samples were studied by XRD with $CuK_\alpha$ radiation using a Philips diffractometer, model PW 1820, having a proportional counter and a graphite monochromator. The measurements were done in the Bragg-Brentano geometry, in the 2Θ range of 10 - 70°, with a step size of 0.02° and a measuring time of 1 s per step. Structural analysis of the samples was done using the Topas Academic V4 software for Rietveld refinement. As illustrated in Fig. 1 the XRD patterns confirmed the fully crystallized state of all samples studied. The magnetic susceptibility of the crystallized alloys $\chi_x$, was measured in the same way as $\chi_a$. The error in the absolute values of $\chi_a$ and $\chi_x$ was about ±2%.

Since both, $\chi_a$ and $\chi_x$ showed very weak dependence on temperature [20, 34, 36], in the following analysis we will use their room temperature values. As already noted [1], the magnetic susceptibility and other properties of MGs which are directly related to the ES (such as the coefficient of the linear term in LTSH and $N(E_F)$) are rather insensitive to the actual quenching conditions (e.g., [20]) which is beneficial for their application as a criterion for GFA.

## 3. Results and Discussion

In Fig. 1 we show XRD patterns of selected Cu-Hf, Cu-Ti and Cu-Zr crystallized samples. As noted earlier [1, 38], the crystallization of Cu-Hf MGs becomes more complex with increasing Cu content (which may contribute to the enhancement of GFA [17, 20, 21] observed at higher Cu-contents [1,16, 37]). In particular, the XRD pattern of the $Cu_{40}Hf_{60}$ sample shows almost pure body centered tetragonal (bct) $CuHf_2$ phase, whereas that of the $Cu_{50}Hf_{50}$ alloy already shows a complex mixture of bct $CuHf_2$ and the orthorombic (o) $Cu_{10}Hf_7$ phase. The crystallization products in all our Cu-Hf alloys were consistent with those observed in previous studies [38] and were also in accord with the CPs appearing in the corresponding composition range of the phase diagram of the Cu-Hf alloy system, for example[39]. The CPs in our crystallized Cu-Zr MGs were also consistent with those observed in previous studies of crystallization in Cu-Zr MGs [34, 40]. In particular, $Cu_{33}Zr_{67}$ MG crystallized directly into the bct $CuZr_2$ compound (Fig. 1) and, as for Cu-Hf MGs, the crystallization became more complex at higher Cu-contents. Similarly, our $Ni_{30}Zr_{70}$ MG crystallized into the bct $NiZr_2$ phase with a small amount of α-Zr phase, which agrees with previous results for the crystallization of Ni-Zr MGs [36, 40].

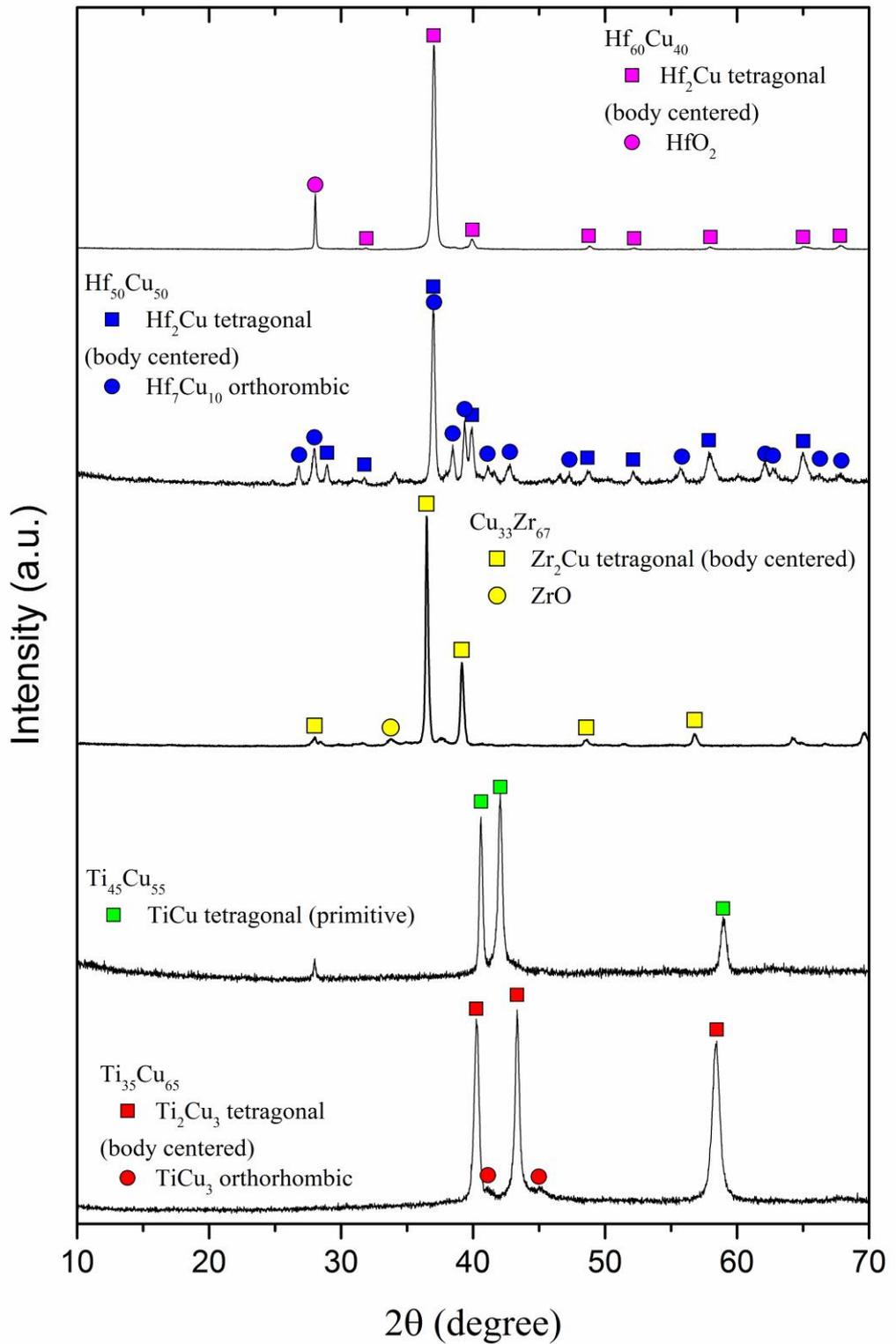

Figure 1. XRD patterns of selected Cu-Hf, Cu-Ti and Cu-Zr crystallized samples. All samples are fully crystallized.

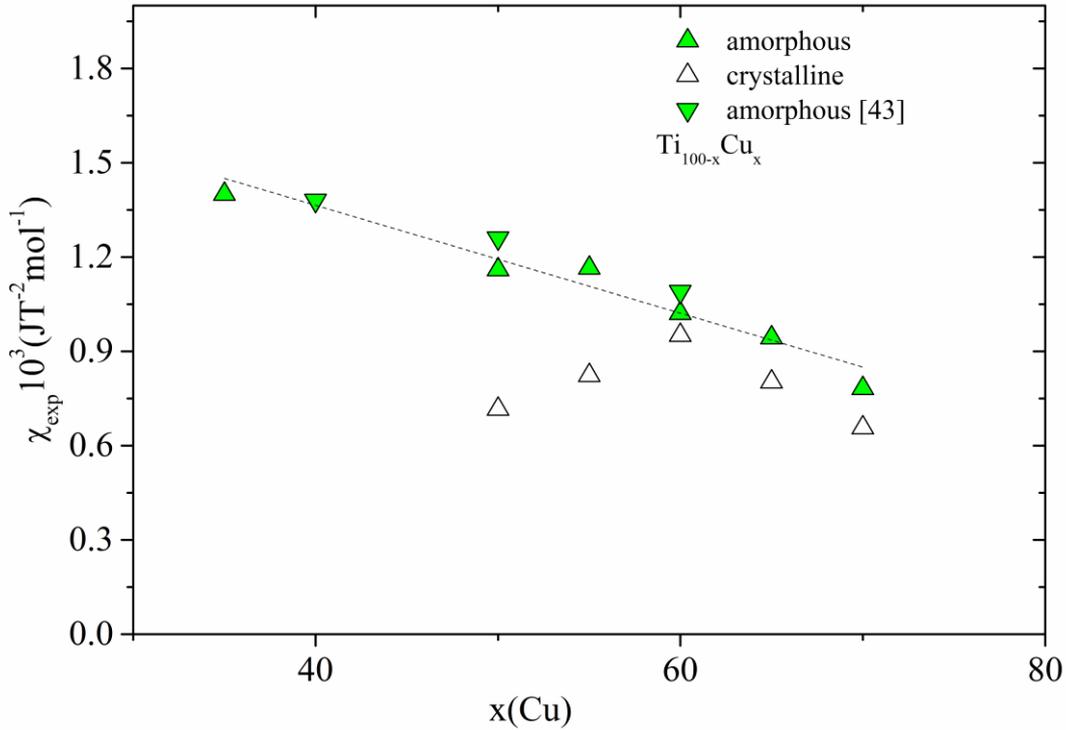

Figure 2. Magnetic susceptibilities of amorphous ($\chi_a$) and crystallized ($\chi_x$) Cu-Ti alloys. Data denoted with reversed triangle are from [43]. Our data (triangles) agree quite well with those from [43]

Although at lower Cu-contents crystallization in all Cu-Ti, Zr, Hf MGs starts with precipitation of the bct $CuTE_2$ phase [34, 36, 38, 40, 41], the overall crystallization pattern in Cu-Ti MGs is somewhat different from that observed in Cu-Zr, Hf MGs [34, 38, 40]. Note that the phase diagram of Cu-Ti system is also different from those for Cu-Zr and Cu-Hf alloys [39]. As shown in Fig. 1 the main crystallization product for a $Cu_{55}Ti_{45}$ alloy is the tetragonal CuTi phase, which contrasts sharply with more complex crystallization patterns in $Cu_{50}Hf_{50}$ and other roughly equiatomic Cu-Hf and Cu-Zr MGs [34, 38, 40]. As illustrated in Fig. 1, for the $Cu_{65}Ti_{35}$ alloy at high Cu-contents in addition to equilibrium phases [39], such as bct $Cu_3Ti_2$, a non-equilibrium orthorhombic $Cu_3Ti$ phase starts to appear which agrees with the results of previous studies of the crystallization behaviour in Cu-Ti MGs [41]. These differences in crystallization behaviour, and thermal stability of Cu-Ti MGs with respect to those of Cu-Zr, Hf MGs may also contribute to lower GFA in the former alloy system [20]. Due to the very high enthalpies of formation of zirconium and hafnium oxides it is difficult to avoid a small amount of surface oxidation during crystallization of TE-TL ribbons (Fig. 1). Because of this we checked the influence of a small amount of surface oxidation on the magnetic susceptibility of our ribbons, by comparing the results for samples showing different amounts of oxide phases in their XRD patterns, and found it to be negligible.

In Fig. 2 we compare the variations with composition of the room temperature magnetic susceptibilities of amorphous $\chi_a$ and crystallized $\chi_x$ Cu-Ti alloys. In spite of the rather complex structure of the magnetic susceptibility in TE-TL MGs [20, 35, 42] the nearly linear decrease of $\chi_a$ with x is qualitatively the same as that of $N(E_F)$ [20] and reflects the approximately linear decrease of the orbital diamagnetism and the Pauli paramagnetism of the d-band with Cu content. As seen in Fig. 2, our results for $\chi_a$ of Cu-Ti MGs agree quite well with the corresponding literature data for the same alloy system [43]. As noted earlier [20, 35] the linear variation of $\chi_a$ with composition in all Cu-Ti, Zr, Hf MGs does not indicate any variation of GFA with composition or any compositions suitable for the formation of BMG in Cu-Zr, Hf MGs [1, 20, 35].

As could be expected from the fact that different crystalline structures occur at different compositions in crystallized Cu-Ti alloys (Fig. 1 and [41]), $\chi_x$ exhibits non-monotonic variation with x showing a maximum for x ≈ 60. We note that in favourable cases one can assign the contributions to magnetic susceptibility from different crystalline phases present in a given alloy, for example Nomura et al in [42]. However, the variations of $\chi_a$ and $\chi_x$ with Cu content in Cu-Ti alloys are qualitatively the same as

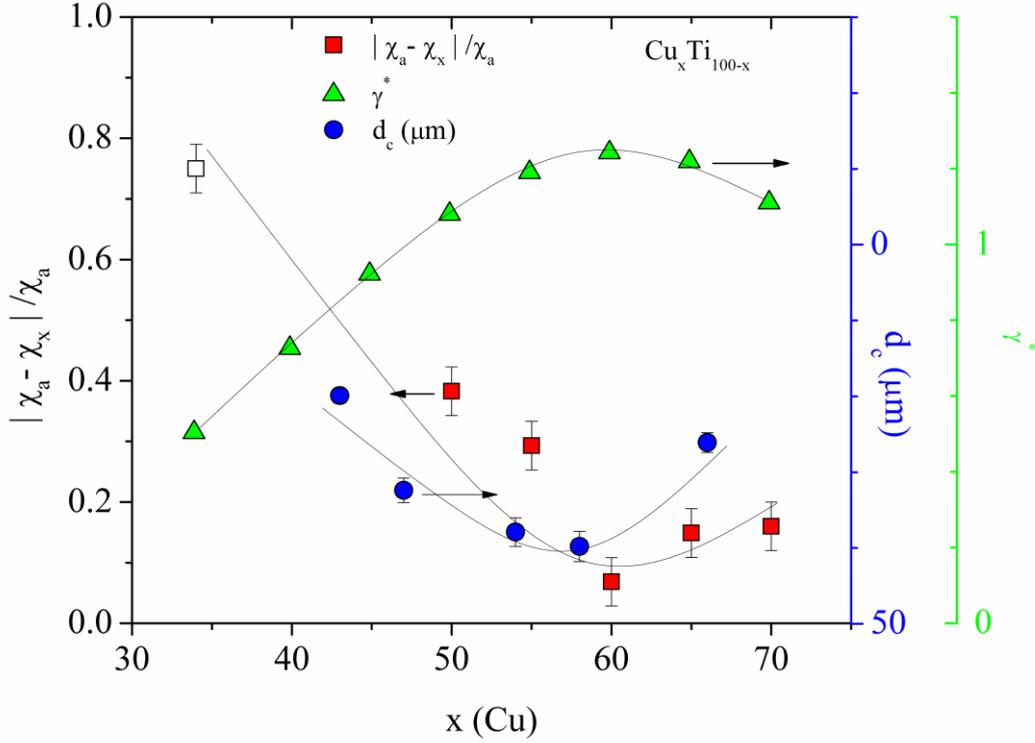

Figure 3. Left scale : ($\Delta\chi_{exp}/\chi_a$) of Cu-Ti alloys vs. concentration x(Cu) . Open symbol is from [45]. Right scale: $d_c$ [44] vs. concentration x(Cu) (first right scale), and $\gamma^*$ vs. x (second right scale). Lines are guides for the eyes.

those in Cu-Hf and Cu-Zr alloys [1, 34]. Therefore, as in the case of Cu-Hf and Cu-Zr alloys [1], it seems advantageous to compare the variation of the fractional change in magnetic susceptibility upon crystallization ($\Delta\chi_{exp}/\chi_a$) with composition in our Cu-Ti alloys with variations of the parameters related to GFA in this system [20, 44]. As seen in Fig. 3, and as found previously for Cu-Hf and Cu-Zr alloys [1], the variation of ($\Delta\chi_{exp}/\chi_a$) with Cu-content in Cu-Ti alloys agrees reasonably well with variations of $d_c$ [44] and $\gamma^* = \Delta H_{amor}/(\Delta H_{inter}-\Delta H_{amor})$, where $\Delta H_{amor}$ and $\Delta H_{inter}$ are the formation enthalpies of glasses and intermetallic compounds [16], respectively, which reflect the GFA in these alloys [20, 44]. In particular, $|\Delta\chi_{exp}/\chi_a|$, like $\gamma^*$ and $d_c$, suggests the largest GFA around x = 60. Further, as expected from rather poor GFA in Cu-Ti alloys [20, 41,

44] the values of both $\gamma^*$ and $d_c$ are quite small and their small magnitude causes considerable uncertainty in the actual values of $d_c$ [44].

There are two important points concerning the data and analysis presented in Fig. 3. First, the critical thickness and therefore the GFA of the ribbons with roughly equiatomic compositions is sizable in spite of the fact that the alloys in this composition range crystallize polymorphously into a single, or nearly single,phase CuTi (Fig. 1, [41, 44]). Thus, in this concentration range, the effects of electronic structure on GFA seem to overcome those associated with kinetic constraints related to phase separation [17, 46]. Second, the advantage of the comparison in Fig. 3 with respect to that for Cu-Hf alloys in Fig. 2 of [1] is that GFA is represented by experimental $d_c$ values which are directly related to GFA, whereas in [1] the GFA was represented with an empirical $T_{rg}$ criterion which does not reflect GFA well in a number of alloys [14], including the Cu-Ti alloy system [20, 44]. Therefore, for a proper assessment of the validity of our, or any other, empirical criterion for GFA, the variation of GFA predicted by a given criterion should be compared with that of $d_c$ or $R_c$, whenever possible.

Accordingly, in Fig. 4 we compare the variation of $\Delta\chi_{exp}$ in Cu-Hf alloys [1] with the corresponding variations of $d_c$ [37] and $\gamma^*$ [20]. (In Ref. 37 critical thicknesses of Hf-Cu ribbons are denoted by $x_c$, but here for simplicity we use the same symbol $d_c$ both for Hf-Cu and Ti-Cu ribbons.) We note that the variations of $\Delta\chi_{exp}$ and $d_c$ with composition are very similar, which provides strong support for the applicability of the $\Delta\chi_{exp}$ ($\Delta ES$) criterion for determination of GFA in Cu-Hf alloys. In particular, small $\Delta\chi_{exp}$ corresponds to large $d_c$, thus to high GFA. The variation of $\gamma^*$, calculated in [20], is also fairly similar to those of $d_c$ and $\Delta\chi_{exp}$. However, our $\gamma^*$ seems to underestimate the GFA of alloys around equiatomic composition and predicts maximum GFA close to x = 65.

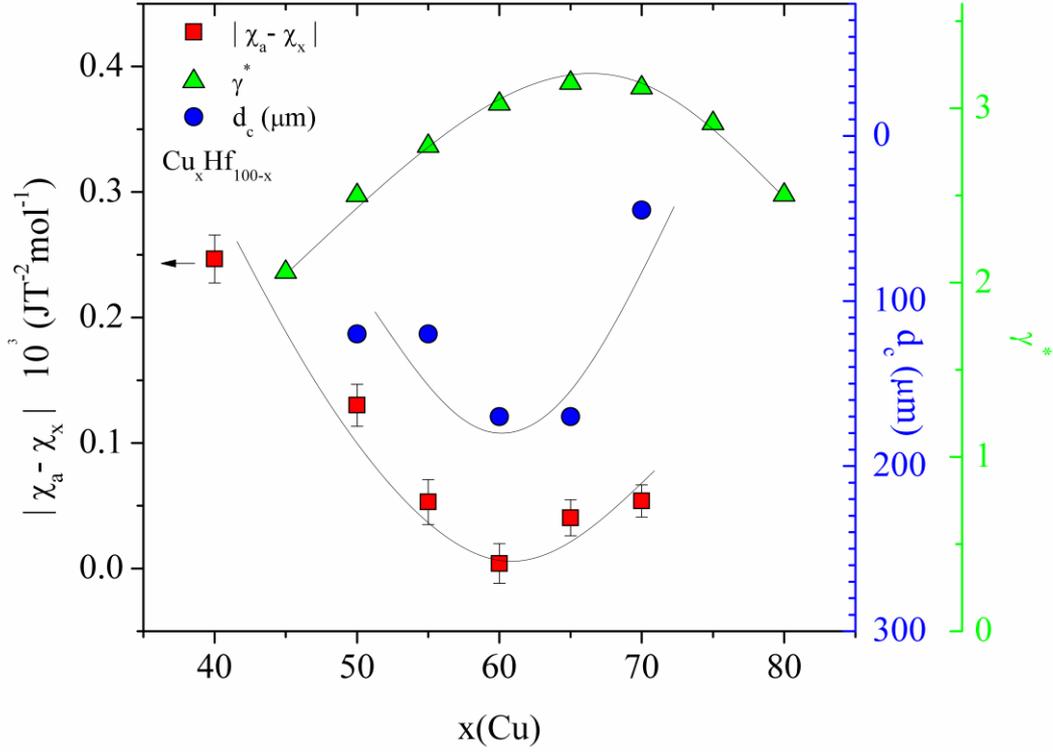

Figure 4. Left scale: $|\chi_a-\chi_x|$ of Cu-Hf alloys vs. concentration x(Cu) . Right scale: $d_c$ [37] vs. concentration x(Cu) (first right scale), and $\gamma^*$ vs. x (second right scale). Note similar variations of $|\chi_a-\chi_x|$ and $d_c$ with x.

We note however that the variation of $\gamma^*$ with composition is very sensitive to which intermetallic compounds are included in calculation of $\gamma^*$ [16]. Thus the choice of intermetallic compounds from Cu-Hf system in [20] may have affected the agreement between $\gamma^*$ and $d_c$ , i.e.GFA, in Fig. 4. Further, as expected from the high GFA of some Cu-Hf alloys [1, 16, 20, 37], the maximum values of $d_c$ and $\gamma^*$ in Fig. 4 are about four and three times larger than the corresponding values for Cu-Ti alloys (Fig. 3). Taken together the results presented in Figs. 3 and 4 show quite clearly that for both Cu-Ti and Cu-Hf alloy systems, in spite of their very different GFAs, a smaller $\Delta\chi_{exp}$, and therefore probably a small change in the average ES upon crystallization of the MG, indicates better GFA. As emphasized in [1], we do not exlude the possible influence of kinetic factors [17, 20, 22, 24] on the actual magnitude of $d_c$ or GFA in a particular alloy, but we believe that the contributions associated with ES dominate the overall variation of GFA with composition in these alloys.

Encouraged by the good correlation between our results for $\Delta\chi_{exp}$ and GFA($d_c$) in Cu-Ti and Cu-Hf alloys we next focus our attention on the corresponding $\Delta\chi_{exp}$ and $\Delta$ES results for a wide range of Cu-Zr [33, 34] and Ni-Zr [36] alloys. GFA in Cu-Zr alloys is particularly interesting, because in these binary alloys BMGs form over an unusually broad composition range [16, 17, 47]. Because of this, a massive search for the origin of BMG formation started immediately upon their discovery [16, 17, 20, 47] and is still continuing with variable success , for example [48]. A preliminary comparison of the variations of $\Delta\chi_{exp}/\chi_a$ with composition in Cu-Hf and Cu-Zr [34] alloys was previously reported [1]. Our present goals are to find out whether we can reproduce the literature results for Cu-Zr alloy system, fill the gaps in data for Cu-Zr [34] alloys and also to seek a more direct relationship between a change in ES upon crystallization [31, 33] and GFA in these and possibly other TL-TE alloy systems.

In Fig. 5 we compare the variations of $\Delta\chi_{exp}/\chi_a$, $d_c$ [47] and $\Delta N_0/N_0(E_F)_a$ ($\Delta N_0 = |N_0(E_F)_a - N_0(E_F)_x|$), with composition in the Cu-Zr alloy system, where $N_0(E_F)_a$ and $N_0(E_F)_x$ denote bare electronic densities of states (DOS) at the Fermi level ($E_F$) of the amorphous and crystallized alloy [33]) respectively. Since $d_c$s in [47] covered only the BMG forming composition we added our estimates for $Zr_{60}Cu_{40}$ and $Zr_{30}Cu_{70}$ alloys, which are just outside this range, to Fig. 5. As described in [37] we multiplied results for these ribbons with a factor of four in order to estimate those for corresponding rods. We note that our $\Delta\chi_{exp}/\chi_a$ data for $Cu_xZr_{100-x}$ alloys with x = 33, 55 and 70, fit in very well with results from [34]. In particular, our result for x = 33 is within the experimental error the same as that in [34], whereas the result for x = 70 provides important information about a rapid increase of $\Delta\chi_{exp}/\chi_a$, and thus decrease of GFA, above x = 65 which was not available in the literature [33, 34]. Further, our result for x = 55 shifts the minimum of $\Delta\chi_{exp}/\chi_a$ closer to the main maximum of $d_c$ at x = 50 [47]. In order to make the correlation between GFA and ΔES more clear, in our previous report [1] we compared the literature results for $\Delta\chi_{exp}/\chi_a$ [34] with those for $\Delta\gamma/\gamma$ [33] which is equal to the fractional change in the dressed DOS at $E_F$, $N(E_F)$. Here $\Delta\gamma = \gamma_a-\gamma_x$, where $\gamma_a$ and $\gamma_x$ are the coefficients of the linear term in the low temperature specific heat of the amorphous and corresponding crystallized alloys, respectively. The variations of $\Delta\chi_{exp}/\chi_a$ and $\Delta\gamma/\gamma$ of Cu-Zr alloys were qualitatively very similar and both quantities showed shallow minima at 60 at% Cu. Since $N(E_F)$ is enhanced with respect to $N_0(E_F)$ by the electron-phonon interaction [20] and this enhancement depends on composition in Cu-Zr MGs and intermetallic compounds [33], $N(E_F)$ does not represent the ES as well as $N_0(E_F)$ does. Similarly, $\chi_{exp}$ [20, 21] in TL-TE alloys represents less well the ES than $N_0(E_F)$ due to contributions from orbital paramagnetism and Stoner enhancement [1, 20]. Because of this, $\Delta N_0/N_0$ in Fig. 5 provides better insight into the change of ES upon crystallization than $\Delta\gamma/\gamma$ and $\Delta\chi_{exp}/\chi_a$ [1]. Indeed, the variation of $\Delta N_0/N_{0a}$ in Fig. 5 is considerably different from that of $\Delta\chi_{exp}/\chi_a$, and also from that of $\Delta\gamma/\gamma$ in [1], and is qualitatively the same as that of $d_c$ [47] of Cu-Zr MGs. In particular, $\Delta N_0/N_0(E_F)_a$ has a minimum at $x = 50$ which is the composition having the largest $d_c$, thus GFA. As noted in our previous report [1] for the other two Cu-Zr alloys with the best GFAs; those with $x = 56$ and 64 [47] there are neither $\Delta N_0/N_0(E_F)_a$ [33] nor $\Delta\chi_{exp}/\chi_a$ results [34]. Since the $d_c$s for alloys with $x = 56$ and 64 are a little lower than that for an equiatomic alloy, the insertion of their $d_c$s in Fig. 5 would hardly affect this figure. However, the variation in $d_c$ over the entire BMG forming composition range in Cu-Zr alloys is within a factor of two or less [47], so that a possible decrease of $\Delta N_0/N_0(E_F)_a$ associated with maxima of $d_c$ may well be within the error associated with two LTSH measurements[1, 33]. Alternatively, a small additional GFA at compositions of the peaks in $d_c$ may not be associated with $\Delta$ES [48]. However, as noted earlier [1] the measurements of kinetic parameters in Cu-Zr alloy systems reported by Russew et al in [24] showed no correlation between these parameters and GFA. Moreover, Angell´s fragility factor m exhibited a maximum in the composition range with the best GFA, which is opposite to what is expected [4, 22, 24]. However, the kinetic effects may be crucial for the enhancement of GFA in TL-TE alloys on addition of Al [49].

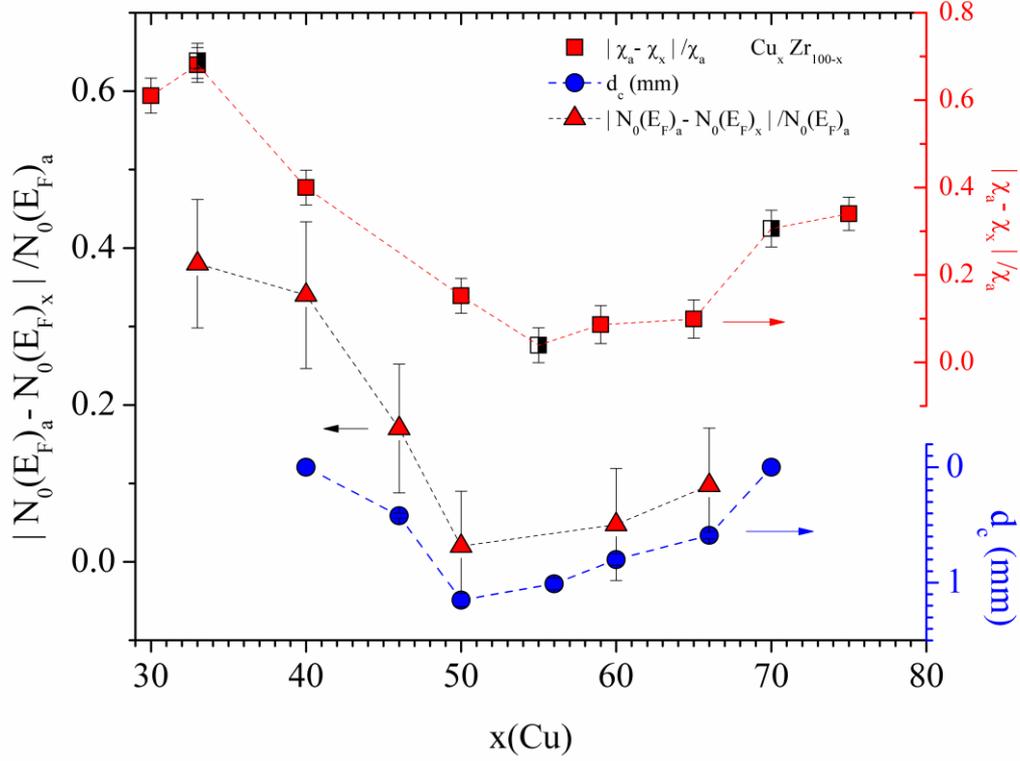

Figure 5. Left scale: $|N_0(E_F)_a - N_0(E_F)_x|/ N_0(E_F)_a$ of Cu-Zr alloys vs. concentration x(Cu). Right scale: $d_c$ [47] vs. concentration x(Cu) (lower right scale), and $|\chi_a-\chi_x|/\chi_a$ vs. x (upper right scale). Half filled symbols are present data. Note maximum of $|\chi_a-\chi_x|/\chi_a$ at $Zr_2Cu$ composition.

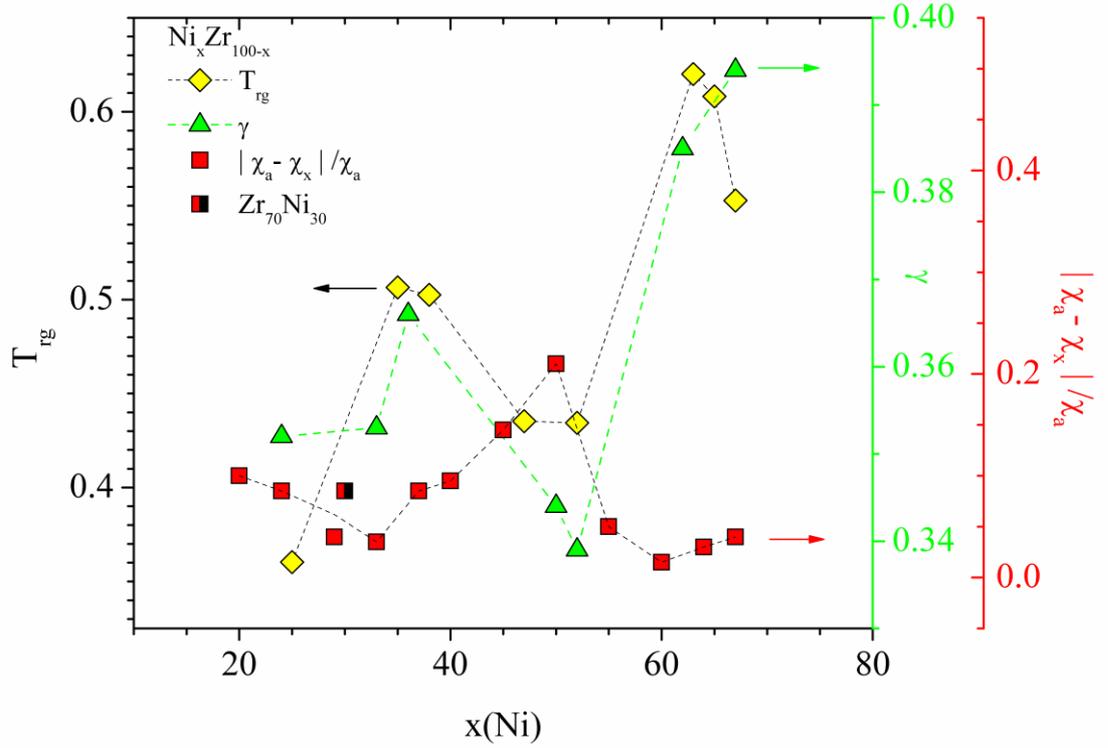

Figure 6. Left scale: $T_{rg}$ [32] of Zr-Ni alloys vs. concentration x(Cu). Right scale: $\gamma_{GFA}$ [32] vs. x (frst right scale) and $|\chi_a-\chi_x|/\chi_a$ vs. x (second right scale). $Zr_{70}Ni_{30}$ alloy is present result. Note maxima of $T_{rg}$ and $\gamma_{GFA}$ (minimum of $|\chi_a-\chi_x|/\chi_a$) close to $Zr_2Ni$ composition.

Simultaneously, $\Delta\chi_{exp}/\chi_a$ shows a sharp maximum at 33.3 at.% Cu where the stable bct $CuZr_2$ compound forms (Fig. 1) directly upon crystallization [34]. The formation of the $CuZr_2$ compound is accompanied with a large decrease of $N_0(E_F)$ with respect to that of corresponding MG as shown in Fig. 2b of [31]. This result [31, 33], together with almost the same variations of $\Delta N_0/N_0(E_F)_a$ and $d_c$ with composition shown in Fig. 5 for Cu-Zr alloys, including the composition range in which BMGs form [47], provides strong support for the proposal that $\Delta ES$ plays a dominant role in GFA of TL-TE alloys.

Next we search for a possible influence of ES on GFA in Ni-Zr alloys. These alloys are particularly interesting because in spite of a wide glass forming range, similar to that in Cu-Zr alloys [34, 40], and similar atomic size and chemical properties of Cu and Ni atoms, Ni-Zr alloys have much lower GFA [36, 40, 44, 50] than Cu-Zr alloys [16, 34, 40, 47]. The origin of this seemingly paradoxical behaviour has been intensely studied over past decades [32, 40, 50-52]. Since there are no data for $d_c$ of Ni-Zr MGs and there is only a single result for $R_c$ of $Ni_{38}Zr_{62}$ alloy [30] in Fig. 6 we compare the variation of $\Delta\chi_{exp}/\chi_a$ [36], which includes also our result for the $Ni_{30}Zr_{70}$ alloy, with those for $T_{rg}$ and $\gamma_{GFA}$ [32] of the same alloy system. There is apparently very good agreement between the variation of these three quantities as shown in Fig. 6 and all three show three extrema at the same compositions. In particular, $\Delta\chi_{exp}/\chi_a$ shows minima around x = 33 and 63 at % Ni where $T_{rg}$ and $\gamma_{GFA}$, and thus presumably GFA, show sharp maxima. Further, $\Delta\chi_{exp}/\chi_a$ is maximum at 50 at% Ni where the GFA, from $T_{rg}$ and $\gamma_{GFA}$, has a deep minimum [32, 50, 51]. In spite of this complex, non-monotonic variation with composition that is presumably associated with rather strong interaction between Ni and Zr atoms [20, 32, 39, 51] a small $\Delta\chi_{exp}/\chi_a$ corresponds to an enhanced GFA as in other TL-TE alloys (Figs. 3-5). Particularly interesting is the local maximum in GFA (minimum in $\Delta\chi_{exp}/\chi_a$) around the composition $Ni_{33}Zr_{67}$ where MGs crystallize directly into bct $NiZr_2$ compound without any phase separation. The probable origin of this puzzling behaviour shows up rather clearly in Fig. 2a of [31] which shows that $N_0(E_F)$ of $NiZr_2$ is only a little lower than that of the corresponding MG. Because of the very similar viscosities of undercooled liquids in Cu-Zr [24] and Ni-Zr [51] alloys it seems quite likely that kinetic effects only affect GFA a little in Ni-Zr alloys. Thus, as in the other TL-TE alloys discussed in the text above, the variation of GFA in Ni-Zr alloys also seems to be dominated by band structure effects.

However, the overall GFA in the Ni-Zr alloys is much smaller than that in the Cu-Zr alloys which may result from smaller difference in local atomic arrangements of glassy and competing CP(s) in a former alloy system [52]. Indeed, in alloy systems in which there is substantial difference in local atomic arrangements of glassy and competing CP(s) and as is the case in Cu-Zr alloys [52], the GFA is likely to be enhanced [1, 3]. The rather strong interatomic interactions in Ni-Zr alloys, associated with strong chemical short range order (CSRO) in corresponding MGs, for example. Bakonyi [23] are likely to diminish the difference in atomic arrangements of the glassy and competing CP(s) in this alloy system.

As noted earlier [1] due to some common properties of binary TL-TE MGs [4, 20, 33-38] and multicomponent transition metal-metal type BMGs [21, 23] the above correlation between the ES and GFA is likely to apply to these BMGs, too. Indeed, recent measurements of LTSH in $Zr_{52.5}Cu_{17.9}Ni_{14.6}Al_{10}Ti_5$ and $Cu_{60}Zr_{20}Hf_{10}Ti_{10}$ BMGs showed very small change in ES upon primary crystallization [53]. Further, in the $Zr_{41}Ti_{14}Cu_{12.5}Ni_{10}Be_{22.5}$ BMG the difference between the density and bulk modulus which is directly related to the ES in the glassy and primary crystallized states was 1 and 3% respectively [54]. However, we expect that these ES based criteria for the GFA apply to all nonmagnetic metal-metal type MGs and BMGs, including those between normal and noble metals [4, 7, 8]. Moreover, in nonsuperconducting alloys of normal metals the magnetic susceptibility and the coefficient of a linear contribution to LTSH should provide a quite accurate description of ES without needing to resort to $N_0(E_F)$. However, as noted in the Introduction, in alloys other than those between early and late transition metals, the DOS at $E_F$ may not be an adequate representation of the actual ES, therefore the comparison of ES in glassy and corresponding crystalline alloys may become more complex.

## 4. Conclusion

The results shown in Figs. 3, 4, 5 and 6 support a plausible close connection between similar electronic band structure (ES) in the glassy and primary crystallized states and glass forming ability (GFA) in four representative alloy systems composed from early and late transition metals, i.e. Cu-Ti, Zr, Hf and Ni-Zr alloys. In particular, for all these alloys, irrespective of their actual GFA and crystallization patterns, the fractional changes in magnetic susceptibility ($\Delta\chi_{exp}/\chi_a$) and/or bare density of states at the Fermi level ($\Delta N_0/N_0(E_F)_a$) follow the same pattern: a small $\Delta\chi_{exp}/\chi_a$ and/or $\Delta N_0/N_0(E_F)_a$ reflect an enhanced GFA, including the formation of bulk metallic glasses (BMG) for some compositions in Cu-Zr, Hf alloys [16, 47], and viceversa. This correlation shows up particularly well in Cu-Zr alloys where $\Delta N_0/N_0(E_F)_a$ shows practically the same variation as the critical casting thickness (Fig. 5). Further, the connection of ES and GFA provides a simple explanation for the paradoxical local maximum of GFA in Ni-Zr alloys at $Ni_{33}Zr_{67}$ composition (Fig. 6 and [31]). Thus, $\Delta\chi_{exp}/\chi_a$ or $\Delta N_0/N_0(E_F)_a$ seem to provide reliable criteria for GFA in binary alloys of early and late transition metals. As discussed in some detail in our previous preliminary report [1] the combination of this criterion with other research methods which can provide information on the atomic structure and electronic structure of metallic glasses and competing intermetallic compounds would be particularly powerful.

Furthermore due to some common properties of binary metallic glasses of late and early transition metals and metal-metal type multicomponent MGs and BMGs the correlation between a change in ES upon crystallization and GFA will probably apply to these alloys, too [1]. Indeed, the recent discovery that even the properties of amorphous, high entropy alloys containing early and late transition metals [28] follow the same behaviour as corresponding binary alloys [20] provides strong support for this claim.

Acknowledgement

We thank Prof. J . R. Cooper for many useful suggestions, and Drs. I. Bakonyi and L. K. Varga for giving us Cu-Hf samples with x =30 and 40. Our research was supported by the project "IZIP2016" of the Josip Juraj Strossmayer University of Osijek. I. A. Figueroa acknowledges the financial support of UNAM-DGAPA-PAPIIT, project No. IN101016.